\documentstyle[12pt,iopfts,psfig]{ioplppt}
\newcommand{\bea}{\begin{eqnarray}}
\newcommand{\eea}{\end{eqnarray}}        
\newcommand{\be}{\begin{equation}}
\newcommand{\ee}{\end{equation}}
\newcommand{\bt}{\begin{tabular}}      
\newcommand{\et}{\end{tabular}}
\newcommand{\Tra}{{\rm Tr}} 
\newcommand{\no}{\nonumber}
\newcommand{\ovl}{\overline}

\begin{document}
\title{Hypermatter in chiral field theory}

\author{P Papazoglou, D Zschiesche S Schramm, H St\"ocker,  W Greiner}

\address{Institut f\"ur Theoretische Physik, 
Postfach 11 19 32, D-60054 Frankfurt am Main, Germany}

\begin{abstract}
A generalized Lagrangian for the description of hadronic matter based 
on the linear $SU(3)_L \times SU(3)_R$ $\sigma$-model is proposed. 
Besides the baryon octet, the spin-0 and spin-1 nonets, a gluon condensate 
associated with broken scale invariance is incorporated. 
The observed values for the vacuum masses of the baryons and mesons
are reproduced. In mean-field 
approximation, vector and scalar interactions yield a 
saturating nuclear equation of state. Finite nuclei can be 
reasonably described, too. The condensates 
and the effective baryon masses at finite baryon density and 
temperature are discussed. 
\end{abstract}
\section{Introduction}
Although the underlying theory of strong interactions is known,
there is presently little hope to solve QCD at nonzero baryon density. \\
Instead, we formulate an effective theory based on symmetries
which hopefully reflects the basic features of QCD in a solvable manner. 
Chiral and flavor symmetries, 
as well as scale invariance are used as guidelines to construct and 
to explore a $\sigma$-model type  $SU(3)_L \times SU(3)_R$ Lagrangian 
at finite strangeness, baryon density and temprature. 
\section{Theory}
Numerous attempts exist to apply the idea of chiral symmetry to nuclear physics. 
A convenient approach is to take the linear $\sigma$ model introduced 
by Gell-Mann and Levy. 
However, wenn calculating the equation of state for nuclear matter 
with this model, bifurcations and multiple solutions lead to an unphysical 
equation of state. 
This suggests that an important physics ingredient is missing. We 
show that the inclusion of (broken) scale invariance allows 
for a satisfactory description of the hadronic mass spectrum, 
nuclear matter and finite nuclei. How this symmetry is incorporated 
in an effective Lagrangian at mean-field level will be discussed in 
subsection \ref{chiralpot}.
\subsection{Nonlinear Realisation of Chiral Symmetry}
We construct a chiral Lagrangian adopting a nonlinear 
realization of chiral symmetry as discussed in \cite{wein68,cole69,callen69}.
The idea is to use a representation where the heavy particles (i.e. baryons 
and scalar mesons)
transform equally under left and right rotations. 
To accomplish this, it is 
necessary to dress these particles nonlinearly with pseudoscalar mesons. 
The application of this method to our approach has the 
advantage that the interaction between heavy particles is only 
governed by $SU(3)_V$. In addition, pseudoscalar particles 
appear explicitly in the Lagrangian only if derivative 
or explicit symmetry breaking terms are included. 
In the following we will outline the argumentation. 
For a thorough discussion, see \cite{paper3}.\\
Let us write the elementary spinors (=quarks) $q$ which transform in the 
whole chiral space $SU(3)_L \times SU(3)_R$ 
into 'new' quarks $\tilde{q}$ by
\be
\label{nlq}
  q_L(x) = U(x) \tilde q_L(x) \qquad q_R(x) = U^{\dagger}(x) \tilde q_R(x)  
\ee
with the pseudoscalar octet $\pi_a$ arranged in $U(x) = \exp[-i\pi_a 
\lambda^a/2]$. 
From the algebraic composition of mesons in terms of quarks (e.g., spin-0 mesons 
have the form $M=\Sigma+i\Pi \simeq \ovl{q}_L q_R+\ovl{q}_L\gamma_5 q_R$), 
it is 
straightforward to transform from `old' mesons 
$\Sigma$ and $\Pi$ (which transform in the whole $SU(3)_L\times SU(3)_R$ space)
into `new' mesons $X$ and $Y$ (which transform in $SU(3)_{L+R}=SU(3)_V$ space):
\be
  M = \Sigma+i\Pi = U (X+iY)U \quad .
\ee
Here, the parity-even part $X$ is associated with the scalar nonet, whereas 
$Y$ 
is taken to be the pseudoscalar singlet \cite{stoks96}.
All hadron multiplets are expanded in a basis of Gell-Mann matrices, 
e.g., $X=\frac{1}{\sqrt{2}}\sum_{a=0}^8 \xi_a\lambda_a$.  
\subsection{Baryon-meson interaction}

In a similar way, the `old' baryon octet 
$\Psi$ forming the representation (8,1) and (1,8) 
is transformed into a `new' baryon octet  
$B=\frac{1}{\sqrt{2}}\sum_{a=1}^8 b_a\lambda_a$:
\be
 \Psi_L = U B_L U^{\dagger} \qquad \Psi_R = U^{\dagger} B_R U \quad .
\ee
The transformations of the exponential $U$ are known 
\cite{cole69,callen69}, 
\be
     U' = LUV^{\dagger} = VUR^{\dagger}, 
\ee
and by using the transformation properties of the `old' fields 
(which can be deduced once the algebraic composition in terms of 
the chiral quarks is known) 
the `new' baryons $B$ and the `new' scalar mesons $X$  
can be shown to transform in $SU(3)_V$. \\
The pseudoscalars reappear in the transformed model 
as the parameters of the symmetry transformation.  
Therefore,  chiral invariants (without space-time derivatives) are  
independent of the Goldstone bosons. The `new' fields 
allow for invariants which are forbidden for 
the `old' fields by chiral symmetry: 
The invariant linear interaction terms of baryons with 
scalar mesons are
\be
{\cal L}_{BX} = g_8^S \Tra\left(\alpha[\ovl{B}BX]_F+ (1-\alpha)  [\ovl{B} B X]_D 
\right)
+ g_1^S \Tra\cdot(\ovl{B} B)\Tra X  \, ,  
\ee
with antisymmetric $[\ovl{B}BX]_F:=\Tra(\ovl{B}BX-\ovl{B}XB)$ and symmetric 
$[\ovl{B}BX]_D:= \Tra(\ovl{B}BX+\ovl{B}XB) - \frac{2}{3} 
\Tra (\ovl{B} B) \Tra M$  invariants. 
The masses of the 
whole baryon multiplet are generated spontaneously by the vacuum 
expectation values (VEV) of only {\it two} meson condensates: From the spin-0 fields 
only the VEV of the components proportional to $\lambda_0$ and the 
hypercharge $Y \sim \lambda_8$ are nonvanishing, and the vacuum expectation 
value $\langle M \rangle$ reduces to: 
\begin{displaymath}
\langle M \rangle\frac {1}{\sqrt{2}}(\xi_0 \lambda_0+\xi_8 \lambda_8)
\equiv \mbox{diag } (\frac{\sigma}{\sqrt{2}} \,, \frac{\sigma}{\sqrt{2}} \,, 
\zeta ) \quad , 
\end{displaymath}
in order to preserve parity invariance and 
assuming, for simplicity, $SU(2)$ symmetry of the vacuum.
Hence, the baryon masses read: 
\bea
\label{bmassen2}
 m_N  &=& m_0 -\frac{1}{3}g_8^S(4\alpha-1)(\sqrt 2\zeta-\sigma) \\ \no
 m_{\Lambda}&=& m_0-\frac{2}{3}g_8^S(\alpha-1)(\sqrt 2\zeta-\sigma) \\ \no
 m_{\Sigma} &=& m_0+\frac{2}{3}g_8^S(\alpha-1)(\sqrt 2\zeta-\sigma)  \\ \no
 m_{\Xi}    &=& m_0+\frac{1}{3}g_8^S(2\alpha+1)(\sqrt 2 \zeta-\sigma) \no
\eea
with $m_0=g_1^S(\sqrt{2} \sigma+\zeta)/\sqrt{3}$.
The three parameters $g_1^S$, $g_8^S$ and $\alpha$ can be used to fit the baryon masses 
to their experimental values (table \ref{parameter}). 
Then, no additional explicit symmetry breaking term is needed.
For $\zeta=\sigma/\sqrt{2}$ (i.e. 
$f_{\pi}=f_K$), the masses are degenerate, and the vacuum is $SU(3)_V$ 
invariant.\\
Although the construction of invariants is only governed by $SU(3)_V$, 
relations following from chiral symmetry as 
PCAC and the Goldberger-Treiman relation are incorporated. The model also 
allows to predict the masses of the meson nonet at zero and finite density.\\
The interaction of the vector meson nonet 
$V_{\mu}=\frac{1}{\sqrt{2}}\sum_{i=0}^8 v_{\mu}^i\lambda_i$
and the axial vector meson nonet 
$A_{\mu}=\frac{1}{\sqrt{2}}\sum_{i=0}^8 a_{\mu}^i\lambda_i$
with baryons reads:
\be
      {\cal L}_{B V} = g_{8}^V  \Tra( \ovl{B} \gamma^{\mu}[V_{\mu},B]
  +\ovl{B} \gamma^{\mu}\{A_{\mu}\gamma_5,B\})
  + g_1^V \Tra(\ovl{B} B) \gamma^{\mu}\Tra (V_{\mu}+A_{\mu}).
\ee
In the mean-field treatment, the axial mesons have a zero VEV. 
 The relevant fields in the SU(2) invariant vacuum, $v^0_{\mu}$ and 
$v^8_{\mu}$, are taken to have the ideal mixing angle 
$\sin \theta_v =\frac{1}{\sqrt{3}}$, yielding
\bea
\label{holz3.36}
  \phi_{\mu}   &=& v_{\mu}^8 \cos \theta_v - v_{\mu}^0 \sin \theta_v =
  \frac{1}{\sqrt{3}} (\sqrt{2} v^0_{\mu}+v_8^{\mu})\\ \no
  \omega_{\mu} &=& v_{\mu}^8 \sin \theta_v + v_{\mu}^0 \cos \theta_v 
 =\frac{1}{\sqrt{3}} (v^0_{\mu}- \sqrt{2} v_8^{\mu}) \quad .  
\eea
For $g_1^V=g_8^V$, the strange vector field $\phi_{\mu} \sim  
\ovl{s}\gamma_{\mu} s $ 
does not couple to the nucleon, and the coupling constant $g_{N\omega}$
is fixed to the binding energy of nuclera matter. 
The remaining 
couplings to the strange baryons are then determined by symmetry relations:
\be
\label{quarkcoupling}
 g_{\Lambda \omega} = g_{\Sigma \omega} = 2 g_{\Xi \omega} = \frac{2}{3} 
 g_{N \omega}=2 g_8^V \qquad 
 g_{\Lambda \phi} = g_{\Sigma  \phi} = \frac{g_{\Xi \phi}}{2} = 
   \frac{\sqrt{2}}{3} g_{N \omega}  \quad , 
\ee                              
where their relative values are related to the additive quark model. 
For isospin asymmetric matter, the $\rho$-field has to be taken 
into account. Its coupling constants are also determined by SU(3) 
symmetry to be
\be
     g_{N\rho}=g_{\Xi \rho}=g_{\Sigma \rho}/2=g_{N\omega}/3\quad g_{\Lambda \rho}=0.
\ee

\subsection{Chirally invariant potential}
\label{chiralpot}
The chirally invariant potential includes the mass terms for mesons, their 
self-interaction and the dilaton potential for the breaking of scale 
symmetry. 
For the spin-0 mesonic potential we take all independent combinations of 
mesonic self-interaction terms up to fourth order 
\bea 
\label{mm-pot}
\fl {\cal V}_0 =   \frac{ 1 }{ 2 } k_0 \chi^2 \Tra M^{\dagger} M 
     - k_1 (\Tra M^{\dagger} M)^2 - k_2 \Tra (M^{\dagger} M)^2 \\ \no
     -  k_3 \chi ( \det M + \det M^ \dagger ) 
    + k_4 \chi^4 + \frac{1}{4} \chi^4 \ln \frac{ \chi^4 }{ \chi_0^4 } 
-\frac{\delta}{3} \chi^4 \ln \frac{\det M + \det M^{\dagger}}
{2 \det \langle M \rangle} \quad  . 
\eea
The constants $k_i$ are fixed 
by the vacuum masses of the pseudoscalar and scalar mesons, respectively. 
These are determined by calculating the second derivative of the potential 
in the ground state. \\
The quadratic and cubic form of the  interaction is made scale invariant 
by multiplying it with an appropriate power of the dilaton 
field $\chi$. 
Originally, the dilaton field was introduced by Schechter in order
to mimic the trace anomaly of QCD
$\theta_{\mu}^{\mu}= \frac{ \beta_{QCD} }{2 g} {\cal G}_{\mu \nu}^a {\cal 
G}^{\mu \nu}_a$
in an effective Lagrangian at tree level \cite{sche80}. 
The effect\footnote{According to \cite{sche80}, the 
argument 
of the logarithm has to be chirally and parity invariant. This is fulfilled by 
the dilaton which is 
a chiral singlet and a scalar.} of the logarithmic term $ \sim \chi^4 \ln \chi$ is 
to break the scale invariance of the model Lagrangian  
so that the proportionality $\theta_{\mu}^{\mu} \sim \chi^4$ 
holds.   
The comparison of the trace anomaly of 
QCD with that of the effective theory allows 
for the identification of the $\chi$-field with the gluon condensate: 
\be
\theta_{\mu}^{\mu} =  \langle \frac{ \beta_{QCD} }{2 g} {\cal G}_{\mu \nu}^a 
{\cal G}^{\mu \nu}_a \rangle
 \equiv  (1-\delta)\chi^4 \qquad .
\ee
The parameter $\delta$ originates from the second logarithmic term with the 
chiral and parity invariant combination $\sim \det M+\det M^{\dagger}$. 
The term is a SU(3)-extension of the logarithmic term proportional to 
$\chi^4 \ln (\sigma^2+ \pi^2)$ introduced in \cite{heid94}. An orientation 
for the value of $\delta$ may be taken from $\beta_{QCD}$ at one loop level, 
which suggests 
the value $\delta=2/11$ for three flavors and three colors. This value 
gives the order of magnitude about which the parameter $\delta$ will be varied.\\
For the spin-1 mesons a mass term is needed. The simplest, scale invariant 
form
\be
\label{vecfree}
{\cal L}_{vec}^1= \frac{1}{2} m_V^2 \frac{\chi^2}{\chi_0^2} \Tra (V_{\mu} V^{\mu}
+ A_{\mu} A^{\mu}) 
\ee
($\chi_0$ is the VEV of the dilaton field) implies a mass degeneracy for the meson nonet. To split the masses one can 
add the chiral invariant \cite{gasi69}
\be
\label{lvecren}
{\cal L}_{vec}^2 = \frac{1}{8} \mu \Tra[(F_{\mu \nu} +G_{\mu \nu})^2 
M^{\dagger} M +(F_{\mu \nu}- G_{\mu \nu})^2 M^{\dagger} M] \quad ,
\ee  
where $F_{\mu \nu}$  and $F_{\mu \nu}$ are the field-strength tensors 
of the vector-and axial-vector fields, respectively. 
In combination with the kinetic energy term (Eq. \ref{kinetic}), 
one obtains for the vector mesons
\bea
\label{kinren}
 {\cal L}_{vec} = &-&\frac{1}{4} [1-\mu \frac{\sigma^2}{2}] (F_{\rho}^{\mu 
\nu})^2
 -\frac{1}{4} [1-\frac{1}{2} \mu (\frac{\sigma^2}{2}+\zeta^2)] 
   (F_{K^{\ast}}^{\mu \nu})^2 \\ \no
 &-&\frac{1}{4} [1-\mu \frac{\sigma^2}{2}](F_{\omega}^{\mu \nu})^2
 -\frac{1}{4} [1- \mu \zeta^2 ] (F_{\phi}^{\mu \nu})^2
\eea 
Since the coefficients are no longer unity, the vector meson fields have 
to be renormalized, i.e., the new $\omega$-field reads 
$\omega_r = Z_{\omega}^{-1/2} \omega$.
The renormalization constants are the coefficients  in the square 
brackets in front of the kinetic energy terms of Eq. (\ref{kinren}), 
i.e., 
$Z_{\omega}^{-1} = 1-\mu \sigma^2/2$. The mass terms of the vector mesons 
deviate from the mean mass $m_V $ by the renormalization 
factor\footnote{One could split the $\rho-\omega$ mass degeneracy by adding 
a term of the form \cite{gasi69} 
$ (Tr F_{\mu \nu})^2$ to Eq. (\ref{kinren}). 
Since the $\rho-\omega$ mass splitting is small 
($\sim$ 2 \%)
, we will not consider this complication.}
, i.e., 
\be
m_{\omega}^2 = m_{\rho}^2=Z_{\omega} m_V^2 \quad ; \quad   
m_{K^{\ast}}^2 = Z_{K^{\ast}} m_V^2 \quad ; \quad 
m_{\phi}^2 = Z_{\phi} m_V^2  \quad .
\ee
The constant $\mu$ is fixed to give the correct $\omega$-mass. The other 
vector meson masses are displayed in table \ref{parameter}.
 
\subsection{Explicit breaking of chiral symmetry}
The term 
\be
{\cal V}_{SB} = \frac{1}{2}\frac{\chi^2}{\chi_0^2} \Tra f (M+M^{\dagger})
\ee
breaks the chiral symmetry explicitly and makes the pseudoscalar mesons massive. 
It is scaled appropriately to have dimensions equal to that of the 
quark mass term 
$\sim m_q \overline{q} q+m_s \overline{s}s$, which is present in the 
QCD Lagrangian with massive quarks. 
This term leads to a nonvanishing divergence of the axial 
currents. The matrix elements of $f=1/\sqrt{2}(f_0 \lambda_0+f_8\lambda_8)$ 
can be written as a 
function of $m_{\pi}^2 f_{\pi}$ and $m_K^2 f_K$ to satisfy 
the (approximately valid) PCAC relations for the $\pi$- and $K$-mesons.
Then, by  utilizing the equations of motion,  
the VEV of $\sigma$ and $\zeta$ are fixed in terms of $f_{\pi}$ and $f_K$. 
\subsection{Total Lagrangian}
Adding the kinetic energy terms for the fermions and  mesons, 
\be
\label{kinetic}
\fl {\cal L}_{kin} = i \Tra \overline{B} \gamma_{\mu} \partial^{\mu}B 
                + \Tra(\partial_{\mu} M^{\dagger} \partial^{\mu} M)
                +\frac {1}{2} \partial_{\mu} \chi \partial^{\mu} \chi 
                - \frac{ 1 }{ 4 } \Tra(F_{ \mu \nu } F^{\mu \nu })  
                - \frac{ 1 }{ 4 } \Tra(G_{ \mu \nu } G^{\mu \nu }) 
\ee
the general Lagrangian is the sum of the various terms discussed: 
\be
\label{lagrange}
{\cal L} = {\cal L}_{kin}+{\cal L}_{BM}
+{\cal L}_{BV}+{\cal L}_{vec}-{\cal V}_{0}-{\cal V}_{SB} \no \qquad .
\ee 
\subsection{Mean field Lagrangian}
To investigate the phase structure of nuclear matter at finite density  we 
adopt the mean-field approximation (see, e.g., \cite{sero97}). 
In this approximation scheme, the fluctuations around constant vacuum 
expectation values of the 
field operators are neglected.
The fermions are treated as quantum mechanical one-particle operators. 
The derivative terms can be neglected and only the 
time-like component of the vector mesons 
$\omega \equiv \langle \omega_0 \rangle$ and 
$\phi \equiv \langle \phi_0 \rangle$ 
survive as we assume homogeneous and isotropic infinite nuclear
 matter. Additionally, due to 
 parity conservation we have $\langle \pi_i \rangle=0$.
After performing these approximations, the Lagrangian (\ref{lagrange}) 
becomes
\begin{eqnarray*}
{\cal L}_{BM}+{\cal L}_{BV} &=& -\sum_{i} \overline{\psi_{i}}[g_{i 
\omega}\gamma_0 \omega^0  +g_{i \rho}\gamma_0 \tau_3 \rho^0
+g_{i \phi}\gamma_0 \phi^0 +m_i^{\ast} ]\psi_{i} \\ \no
{\cal L}_{vec} &=& \frac{ 1 }{ 2 } m_{\omega}^{2}\frac{\chi^2}{\chi_0^2}\omega^2  
+\frac{ 1 }{ 2 } m_{\rho}^{2}\frac{\chi^2}{\chi_0^2}\rho^2
 + \frac{ 1 }{ 2 }  m_{\phi}^{2}\frac{\chi^2}{\chi_0^2} \phi^2\\
{\cal V}_0 &=& \frac{ 1 }{ 2 } k_0 \chi^2 (\sigma^2+\zeta^2) 
- k_1 (\sigma^2+\zeta^2)^2 
     - k_2 ( \frac{ \sigma^4}{ 2 } + \zeta^4) 
     - k_3 \chi \sigma^2 \zeta \\ 
&+& k_4 \chi^4 + \frac{1}{4}\chi^4 \ln \frac{ \chi^4 }{ \chi_0^4}
 -\frac{\delta}{3}\ln \frac{\sigma^2\zeta}{\sigma_0^2 \zeta_0} \\ \no
{\cal V}_{SB} &=& (\frac{\chi}{\chi_0})^{2}[m_{\pi}^2 f_{\pi} \sigma 
+ (\sqrt{2}m_K^2 f_K - \frac{ 1 }{ \sqrt{2} } m_{\pi}^2 f_{\pi})\zeta ] \quad , 
%
\end{eqnarray*}
with 
the effective baryon masses $m_i^{\ast}$, which 
are given in Eqs. \ref{bmassen2}.
\subsubsection{Grand canonical ensemble}
The values for the fields at a given chemical potential $\mu$
are determined by minimizing the
thermodynamical potential of the grand canonical, which 
reads: 
\be
   \frac{\Omega}{V}= -{\cal L}_{vec} + {\cal V}_0 + {\cal V}_{SB}
-{\cal V}_{vac}- \sum_i \frac{\gamma_i }{(2 \pi)^3}  
\int d^3k [E^{\ast}_i(k)-\mu^{\ast}_i]   
\ee 
The vacuum energy ${\cal V}_{vac}$ (the potential at $\rho=0$) 
has to be subtracted in 
order to get a vanishing vacuum energy. $\gamma_i$ denote the fermionic 
spin-isospin degeneracy factors ($\gamma_N=4$, $\gamma_{\Sigma}=6$, 
$\gamma_{\Lambda}=2$, $\gamma_{\Xi}=4$).
The single particle energies are 
$E^{\ast}_i (k) = \sqrt{ k_i^2+{m_i^*}^2}$ 
and the effective chemical potentials read
$\mu^{\ast}_i = \mu_i-g_{\omega i} \omega-g_{\phi i} \phi$. 
The baryon chemical potentials are realated 
to each other by means of 
the nonstrange chemical potential $\mu_q$ and the strange chemical 
potential $\mu_s$ according to the additive quark model:
\bea
   \mu_N &=&3 \mu_q \qquad \mu_{\Xi}=\mu_q+2\mu_s        \\ \no 
   \mu_{\Lambda}&=&2\mu_q+\mu_s=\mu_{\Sigma}             
\eea
The energy density and the pressure  follow from the Gibbs--Duhem relation, 
$\epsilon = \Omega/V+ \mu_i \rho^i$ and $p= - \Omega/V$. 

\section{Results}
\subsection{Fits to nuclear matter and the hadronic spectrum}
A salient feature of all chiral models are the strong vacuum constraints. 
In the present case they fix  $k_0$, $k_2$ and $k_4$,  in
order to minimize the thermodynamical potential $\Omega$ in vacuum
for given values of the fields $\sigma_0$, $\zeta_0$ and $\chi_0$. 
The parameter $k_3$ is  fixed to the $\eta$-mass m$_{\eta}$. 
There is some freedom to vary parameters, mainly due to 
the unknown mass of the $\sigma$-meson, $m_{\sigma}$, 
which is determined by $k_1$, 
and due to the uncertainty of the value for the 
kaon decay constant $f_{K}$.\\ 
It should be noted that a reasonable nuclear matter fit with low 
compressibility can be found 
(table \ref{parameter}), where $m_{\sigma} \approx$ 500 MeV (The other 
scalar mesons have a mass of about 1 GeV \cite{paper2}). 
This, in the present approach, allows for an interpretation of the $\sigma$-field
as the chiral partner of the $\pi$-field and as the mediator  
of the mid-range attractive force between nucleons, though we believe 
the phenomenon is in reality  
generated through correlated two-pion exchange \cite{sero97}.
\begin{table}
\begin{center}
\bt{|c|c|c|c|} \hline
  $m_{\pi}$(139) & $m_{K}$(495) & $m_{\eta}$(547) & $m_{\eta'}$(958) \\ \hline 
     139.0 & 498.0 & 520.0 & 999.4\\ \hline  \hline 
  m$_{\rho}$(770) & m$_{K^{\ast}}$(892) &  m$_{\omega}$(783)& m$_{\phi}(1020)$ \\ \hline
  783.0 & 857.7  & 783.0 & 1019.5 \\ \hline \hline
 m$_N$(939)  & m$_{\Lambda}$(1115) & $m_{\Sigma}$(1193) & m$_{\Xi}$(1315) \\ \hline   
939.0 & 1117.8 & 1193.1 & 1334.5 \\ \hline \hline
K [MeV]& $\frac{m_N^{\ast}}{m_N}$ &$f_{\pi}$ [MeV]& f$_K$ 
[MeV] \\ \hline 
     278.7  & 0.62 & 93.0  & 122.0  \\ \hline 
   \et
\caption{\label{parameter}Fit to the hadronic mass spectrum and to nuclear matter. The results were obtained with the parameter 
k$_0$=2.3, k$_1$=1.4,  k$_2$=-5.5, and k$_3$=-2.6 .}
\end{center}
\end{table}
\subsection{Finite Nuclei}
If the fields are allowed to be spatial dependent, then derivative 
terms do not vanish and it is 
possible to look for the properties of finite nuclei. We adopted 
the Hartree formalism which is explained in \cite{sero97} 
and included a chiral invariant quartic spin-1 meson self-interaction 
of the form $g_4^4 \Tr \left[ (V_{\mu}+A_{\mu})^4+(V_{\mu}-A_{\mu})^4 
\right]$
to allow for a fine tuning of the model.\\ 
The $SU(3)_L \times SU(3)_R \sigma$ model accounts also for a 
satisfactory description of finite nuclei. This is surprising, 
since most of the parameters are fixed to the vacuum masses, 
in contrast to the Walecka-model, where all the parameters 
are adjusted to the properties of nuclear matter or nuclei.  
Although no fit to nuclei is done (two parameters are 
fixed to nuclear matter), the agreement of the observables 
resulting from the model as compared to the experimental 
values is acceptable (table \ref{nuclei}). 
Although there is no freedom to adjust the coupling constant of nucleons 
to the $\rho$ meson, $g_{N\rho}$, the 
asymmetry energy $a_4$ has the reasonable value $a_4=39.3$ MeV. 
\begin{table}
\begin{center}
\bt{|c|cc|cc|cc|} \hline
  & \multicolumn{2}{c|}{$^{16}O$} & \multicolumn{2}{c|}{$^{40}Ca$} & 
  \multicolumn{2}{c|}{$^{208}Pb$}\\ \hline
$E/A$         &7.53  &7.98 &8.15  &8.55 & 7.65  & 7.86\\ 
$r_{ch}$      &2.62  &2.73 &3.40  &3.48 & 5.48  & 5.50\\ 
$R$           & 2.75 &2.78 &3.81  &3.85 & 6.81  & 6.81\\ 
$\sigma$      & 0.81 &0.84 &0.93  &0.98 & 0.89  & 0.90\\ \hline
 \et
\caption{\label{nuclei}
Bulk properties of nuclei:Prediction (left) and experimental values
 (right) for 
binding energy $E/A$, charge radius $r_{ch}$, surface tension $\sigma$ 
and diffraction radius $R$ of Oxigen ($^{16}O$), Calcium 
($^{40}Ca$) and Lead ($^{208}Pb$).} 
\end{center}
\end{table}
\subsection{Condensates and hadron masses in medium}
\begin{figure}
\psfig{file=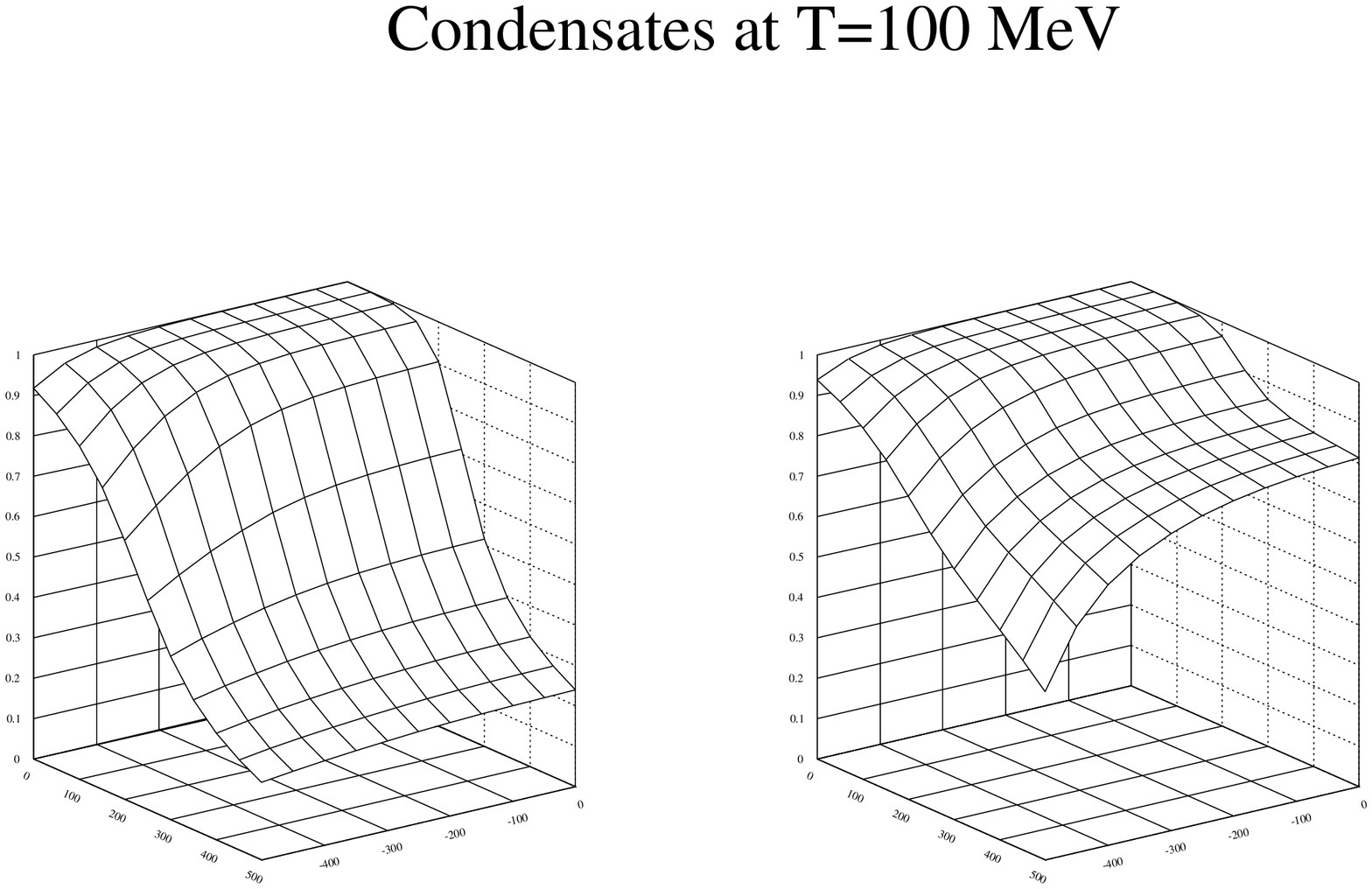,angle=-90,width=180mm}
\caption{\label{felder} Nonstrange and strange condensates versus 
$\mu_q$ and $\mu_s$ at temperature $T=100$ MeV. }
\end{figure}

\begin{figure}
\psfig{file=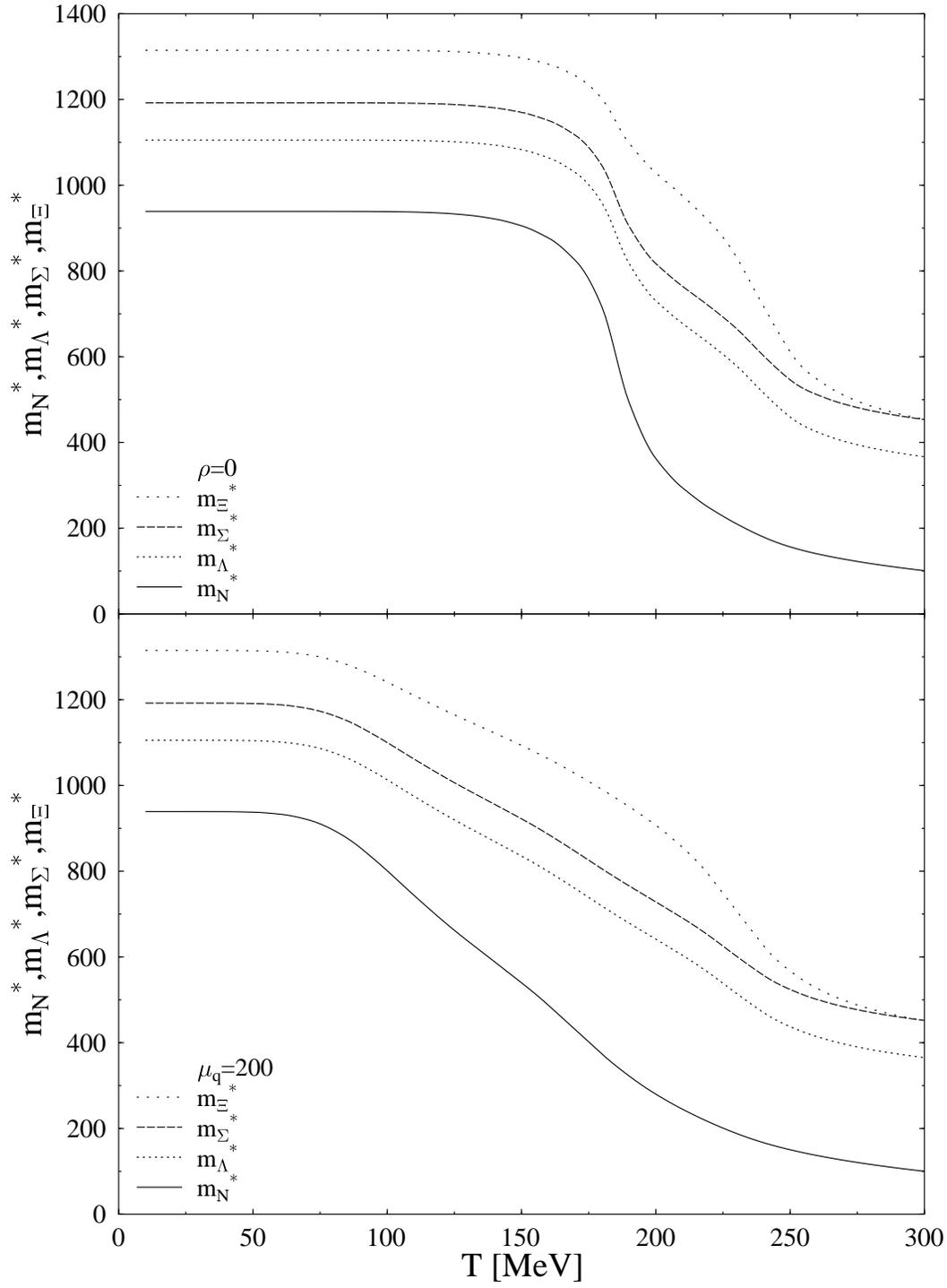,width=150mm}
\caption{\label{massen} Baryon masses as a function of temperature for 
zero (above) and finite (below) chemical potentials.}
\end{figure}

It is instructive to see how the condensates change  
in the hot and dense medium, since these determine how the 
hadron masses change.
The nonstrange condensae $\sigma\sim\langle qq \rangle$ and the strange 
condensate\footnote{For simplicity, we fix 
the gluon condensate at its vacuum expectation value ($\chi=\chi_0$).} 
$\zeta \sim \langle ss\rangle$)
are displayed in Fig. \ref{felder} as a function of $\mu_q$ and 
$\mu_s$ for given temperature $T=100$ MeV. 
The field $\sigma$ decreases rapidly in the $\mu_q$ direction, whereas 
$\zeta \sim \langle ss\rangle$ doesn't change
significantly at this temperature. This is different for $T=200$ MeV.
There, the nonstrange condensate\footnote{The calculation is done 
in the limit of vanishing coupling of the strange condensate $\zeta$ to 
the nucleons, i.e, for $\alpha=1$ and $g_1^S=\sqrt{6}g_8^S$ (see Eq. \ref{bmassen2})}
 has already dropped to $40\%$
of its VEV and shows only a small $\mu_q$ and $\mu_s$ dependence. 
This is in contrast to the strange condensate that 
decreased only to $80\%$ of 
its VEV and reacts rather strongly for increasing $\mu_s$ .\\
The linear dependence of the baryon masses
to the condensates causes a dropping of the effective masses 
(Fig. \ref{massen}) 
with increasing temperature. 
This may lead to an enhanced baryon/antibaryon
production. For higher chemical potentials this decrease is softened
since it sets in earlier but tends towards the same high temperature 
limit, as can be seen comparing the curves for $\mu_q=\mu_s=0$   
and $\mu_q=200$ MeV, $\mu_s=0$. The 'wiggly' behaviour of the hyperon 
masses is due to the decrease of the nonstrange condensate at lower temperatures than 
the $\langle ss\rangle$.\\
We are currently working on the formulation of a chiral
transport theory within the framework of our model
which takes into account the effects of dynamical hadron masses
in a heavy ion collision. These efforts may help to
elucidate possible signatures of a chiral phase transition.

\ack
This work 
was funded in part by Deutsche Forschungsgemeinschaft (DFG), Gesellschaft 
f\"ur Schwerionenforschung (GSI) and 
Bundesministerium f\"ur Bildung und Forschung (BMBF). 

\section*{References}

 \end{document}